\documentclass[12pt]{article}

\begin{document}
\begin{center}
{\Large\bf Gravitational energy-momentum in small regions according
to M{\o}ller's tetrad expression}
\end{center}

\begin{center}
Lau Loi So\footnote{e-mail address: s0242010@cc.ncu.edu.tw,\\
 present address: Department of Physics, Tamkang University,
Damsui 251, Taiwan} and James
M. Nester\footnote{e-mail address: nester@phy.ncu.edu.tw}\ \\
$^{1}$Department of Physics, National Central University,
Chung-Li 32054, Taiwan.\\
$^{2}$Department of Physics and Institute of Astronomy, National
Central University, Chung-Li 32054, Taiwan.
\end{center}

\begin{abstract}  M{\o}ller's tetrad
gravitational energy-momentum ``tensor'' is evaluated for a small
vacuum region using an orthonormal frame adapted to Riemann normal
coordinates. We find that it does satisfy the highly desired
property of being a positive multiple of the Bel-Robinson tensor.
\end{abstract}

\section{Introduction: energy-momentum localization}

The localization of energy-momentum for gravitating systems is still
an outstanding fundamental problem \cite{Sza04}.  The classical
attempts to identify a gravitational energy-momentum density for
Einstein's {\it covariant} theory, general relativity, had all led
to {\it non-covariant}, coordinate system dependent expressions,
generally referred to as pseudotensors (see, e.g.,
\cite{Gol58,Tra62,CNC99,Nes04}). As coordinate systems have no
physical significance, this led to the idea that there was no
physically meaningful gravitational energy-momentum density, and,
moreover, that that is just what we should expect from the
equivalence principle, (see \cite{MTW}, \S 20.4).

Meanwhile, in 1961 M{\o}ller had constructed an energy-momentum
expression which leads to a {\it tensor\/} under coordinate
transformations \cite{Mol61}. This ``tensor'' form is achieved,
however, by introducing an orthonormal frame, a tetrad (aka
vierbein). Thus, while this expression is a tensor with respect to
coordinate transformations, it depends on the local choice of the
orthonormal frame.  More precisely, like many other energy-momentum
expressions, the value it assigns to a spatial region is {\it
quasi-local\/} \cite{Sza04}: it depends on the fields only at the
boundary of the region. The energy-momentum M{\o}ller's ``tensor''
expression assigns to a spacetime region thus depends on an object
which includes non-physical information, namely the choice of tetrad
on the boundary. Nevertheless, largely because of its perceived
advantages for energy-momentum localization, M{\o}ller's tetrad
expression (which, by the way, also admits an interesting
teleparallel representation) has continued to attract interest over
the years (see, e.g., \cite{Nes89,Mal,Maletal99-02,AGP00} and the
works cited therein), even though there is no generally accepted
frame gauge condition.

In certain special cases, however, there is a natural orthonormal
frame; then M{\o}ller's expression yields an unambiguous
energy-momentum. In particular this is so asymptotically---at
spatial infinity.  In that case M{\o}ller's expression (like most
others) works well (see \cite{Des63} for an explicit verification;
moreover Moller's tetrad expression in fact also works well at
future null infinity \cite{Mol64}). This asymptotic success is
actually not at all surprising; having the proper asymptotic
behavior is a relatively weak requirement, for in this weak field
region an expression need only have the proper linear theory limit.

The situation is different in the one other situation where there is
a natural frame---a case which has, to our knowledge, not been
previously investigated for M{\o}ller's expression---namely the
small region limit. In this limit, to zeroth order, one should get
the material energy-momentum density---a quite weak requirement
which follows from the equivalence principle.  On the other hand the
proposed small {\it vacuum\/} region limit is that, to {\it second
order\/}, one gets a positive multiple of the {\it Bel-Robinson
tensor\/} \cite{Gar,DFS99,Sza04} (that would be sufficient to
guarantee that the energy of a small region was {\it positive\/}).
Now this latter requirement is especially interesting as a test of
proposed energy-momentum densities, since it probes the expression
beyond the linear order. It is a strong criterion, capable of
excluding many otherwise acceptable expressions, in particular {\it
none\/} of the classical pseudotensors satisfy this requirement
(although certain artificial combinations of them do
\cite{DFS99,icga7so,So06}).

Here, using Riemann normal coordinates and the associated ``normal''
tetrad, we examine M{\o}ller's expression in the small region limit.
We find that M{\o}ller's expression {\it naturally\/} satisfies this
highly desirable vacuum Bel-Robinson property.

\section{M{\o}ller's energy-momentum tensor}
A gravitational energy-momentum density is easily derived from
Einstein's equations expressed in terms of differential forms:
\begin{equation}
R^\alpha{}_\beta\wedge\eta_\alpha{}^\beta{}_\mu=-2\kappa
T_\mu.\label{eineq}
\end{equation}
Here $\kappa=8\pi G/c^4=8\pi G$ is the coupling constant,
$R^\alpha{}_\beta$ is the curvature 2-form,
$T_\mu=T^\nu{}_\mu\eta_\mu$ is the source energy-momentum 3-form,
and we are using Trautman's convenient dual form basis
$\eta^{\alpha\dots}:=*(\vartheta^\alpha\wedge\dots)$, where
$\vartheta^\alpha$ is the co-frame. The left hand side of
(\ref{eineq}) is just $-2G^\nu{}_\mu\eta_\nu$, the Einstein tensor
expressed as a 3-form. (Our conventions, unless otherwise stated,
follow MTW \cite{MTW}.) Using the definition of the curvature 2-form
in terms of the connection one-form and extracting an exact
differential leads to
\begin{eqnarray} R^\alpha{}_\beta\wedge\eta_\alpha{}^\beta{}_\mu&:=&
(d\Gamma^\alpha{}_\beta+\Gamma^\alpha{}_\gamma\wedge\Gamma^\gamma{}_\beta)\wedge\eta_\alpha{}^\beta{}_\mu\nonumber\\
&\equiv&d(\Gamma^\alpha{}_\beta\wedge\eta_\alpha{}^\beta{}_\mu)
+\Gamma^\alpha{}_\beta\wedge d\eta_\alpha{}^\beta{}_\mu
+\Gamma^\alpha{}_\gamma\wedge\Gamma^\gamma{}_\beta\wedge\eta_\alpha{}^\beta{}_\mu\label{decomp}\nonumber\\
&\equiv&d(\Gamma^\alpha{}_\beta\wedge\eta_\alpha{}^\beta{}_\mu)
+\Gamma^\alpha{}_\beta\wedge\Gamma^\lambda{}_\mu\wedge\eta_\alpha{}^\beta{}_\lambda
-\Gamma^\alpha{}_\gamma\wedge\Gamma^\gamma{}_\beta\wedge\eta_\alpha{}^\beta{}_\mu,
\end{eqnarray}
where we have used $D\eta_\alpha{}^\beta{}_\mu=0$, which follows
since the connection is metric compatible and torsion free.  Using
this expansion we can rewrite the Einstein equation (\ref{eineq}) in
a neat form (which is remarkably similar to the form used by
Einstein when he was still searching for a good gravity theory
\cite{JR06}):
\begin{equation}
dp_\mu=2\kappa{\cal P}_\mu,
\end{equation}
where the energy-momentum flux 2-form is
\begin{equation}
p_\mu:=-\Gamma^\alpha{}_\beta\wedge\eta_\alpha{}^\beta{}_\mu,
\end{equation}
and the  current is the {\it total energy-momentum\/} density
(3-form)
\begin{equation}
{\cal P}_\mu:=t_\mu+T_\mu,
\end{equation}
which ``automatically'' satisfies the current conservation relation
$d{\cal P}_\mu=0$ \cite{Nes04}.  This total energy-momentum current
``complex'' includes the (non-covariant) gravitational
energy-momentum density
\begin{equation}
t_\mu:=(2\kappa)^{-1}\left(\Gamma^\alpha{}_\beta\wedge\Gamma^\lambda{}_\mu\wedge\eta_\alpha{}^\beta{}_\lambda
-\Gamma^\alpha{}_\gamma\wedge\Gamma^\gamma{}_\beta\wedge\eta_\alpha{}^\beta{}_\mu\right).\label{gravemdensity}
\end{equation}
According to this prescription the total energy-momentum within a
region is given by
\begin{equation}
P_\mu(V):=\int_V {\cal P}_\mu = (2\kappa)^{-1}\oint_{\partial
V}p_\mu .
\end{equation}

The volume integral form would lead one to expect that the value
depends on the quantities and choice of frame throughout the region,
but the closed 2-surface integral shows that the value is {\it
quasi-local\/}. The value is  still {\it non-covariant\/}: it
depends on the choice of frame---but, as we have already pointed
out, only on the choice at (and, through the connection, near) the
boundary.

If the frame is {\it holonomic\/} then the flux integrand,
\begin{equation}
p_\mu:=-\Gamma^\alpha{}_\beta\wedge\eta_\alpha{}^\beta{}_\mu\equiv
-\Gamma^\alpha{}_{\beta\gamma}g^{\beta\sigma}\delta^{\tau\rho\gamma}_{\alpha\sigma\mu}{1\over2}
\eta_{\tau\rho},
\end{equation}
 is the Freud
{\it superpotential\/} \cite{Freud} (written as a 2-form) and the
gravitational energy-momentum density is the Einstein {\it
pseudotensor\/} 3-form. If, on the other hand, one chooses the frame
to be orthonormal, then {\it these same formal expressions} (for an
earlier observation of this formal correspondence see \cite{Gol80})
become those proposed by M{\o}ller \cite{Mol61} in 1961 (by the way,
a differential form construction of these expressions virtually the
same as ours was presented some time ago by Wallner \cite{Wal80});
the noteworthy thing is that the tetrad expressions are {\it
tensors\/}---under coordinate transformations. Although they are
completely independent of the choice of coordinates, they do depend
on the choice of tetrad; in this important sense they are
non-covariant. More specifically the energy-momentum values they
determine are quasi-local: they depend on the choice of tetrad, but
only on the choice at and near the boundary.

\section{Riemann normal coordinates and normal tetrads}
To find the energy-momentum within a small region surrounding a
particular point, we look to the 3-form ${\cal P}_\mu$, expanding it
in a power series.  For this purpose we choose Riemann normal
coordinates $x^i$ centered at the selected point.  The
Maclauren-Taylor expansion of the holonomic components of the metric
and connection are well known (see, e.g.~\cite{MTW}, \S 11.6):
\begin{equation}
g_{ij}|_0=\bar g_{ij}, \quad \partial_k g_{ij}|_0=0, \quad
3\partial_{kl}g_{ij}|_0=-R_{ikjl}-R_{iljk},
\end{equation}
\begin{equation}
\Gamma^i{}_{jk}|_0=0,\quad
3\partial_l\Gamma^i{}_{jk}|_0=-R^i{}_{jkl}-R^i{}_{kjl}.
\end{equation}
Here $\bar g_{ij}=\hbox{diag}(-+++)$ is the Minkowski metric.  In
the associated ``normal'' orthonormal frame, the coframe
$\vartheta^\alpha=\vartheta^\alpha{}_kdx^k$ and connection one-form
$\Gamma^\alpha{}_{\beta k}dx^k$ components take closely related
analogous values:
\begin{equation}
\vartheta^\alpha{}_j|_0=\delta^\alpha{}_j,\quad
\partial_k\vartheta^\alpha{}_j|_0=0,\quad
6\partial_{kl}\vartheta^\alpha{}_j|_0=-R^\alpha{}_{kjl}-R^\alpha{}_{ljk},
\end{equation}
\begin{equation}
\Gamma^\alpha{}_{\beta j}|_0=0,\quad 2\partial_k
\Gamma^\alpha{}_{\beta l}|_0=R^\alpha{}_{\beta kl}.
\end{equation}
It is readily verified that these values satisfy, to the appropriate
order, the two relations which transform the metric and connection
coefficients between the holonomic and orthonormal frames:
\begin{equation}
g_{ij}=\bar g_{\alpha\beta}\vartheta^\alpha{}_i\vartheta^\beta{}_j,
\quad  \vartheta^\beta{}_j \Gamma{}^\alpha{}_{\beta
i}=\Gamma{}^k{}_{ji}\vartheta^\alpha{}_k-\partial_i\vartheta^\alpha{}_j.
\end{equation}

\section{Small region values}
Expanding ${\cal P}_\mu$ using Riemann normal coordinates and the
associated normal tetrad gives to zeroth order, unsurprisingly, only
the source energy momentum density---just as it should according to
the {\it equivalence principle\/}.  In vacuum regions ${\cal P}_\mu$
reduces to ${t_\mu}$ (\ref{gravemdensity}), and the leading
non-vanishing value appears at the second order:
\begin{eqnarray}
2\kappa{\cal P}_\mu&=
&\Gamma^{\alpha\beta}\wedge\Gamma^\lambda{}_\mu\wedge\eta_{\alpha\beta\lambda}
-\Gamma^\alpha{}_\gamma\wedge\Gamma^{\gamma\beta}\wedge\eta_{\alpha\beta\mu}\\
&\simeq&\frac{x^lx^m}4\left(R^{\alpha\beta}{}_ {li} R^\lambda{}_{\mu
mj}
-\delta^\lambda_\mu R^\alpha{}_{\gamma li}R^{\gamma\beta}{}_{mj}\right)dx^i\wedge dx^j\wedge\eta_{\alpha\beta\lambda}\nonumber\\
&\simeq&\frac{x^lx^m}4\left(R^{\alpha\beta}{}_ {l\sigma}
R^\lambda{}_{\mu m\delta}
-\delta^\lambda_\mu R^\alpha{}_{\gamma l\sigma}R^{\gamma\beta}{}_{m\delta}\right)\delta^{\nu\sigma\delta}_{\alpha\beta\lambda}\eta_\nu\nonumber\\
 &=&\frac{x^lx^m}4\left(2R_{\mu \lambda m \delta}
R^{\nu\lambda}{}_{l}{}^\sigma
-\frac12\delta^\nu_\mu R^{\gamma\sigma\delta}{}_lR_{\gamma\sigma\delta m}\right)\eta_\nu\\
&=&\frac{x^lx^m}4B^\nu{}_{\mu lm}\eta_\nu,
\end{eqnarray}
proportional to the Bel-Robinson tensor:
\begin{equation}
B_{\alpha\beta\mu\nu}
:=R_{\alpha\lambda\mu\sigma}R_{\beta}{}^\lambda{}_{\nu}{}^{\sigma}
+R_{\alpha\lambda\nu\sigma}R_{\beta}{}^{\lambda}{}_{\mu}{}^{\sigma}
-\frac12g_{\alpha\beta} R^{\gamma\sigma\delta}{}_\mu
R_{\gamma\sigma\delta \nu}.
\end{equation}
In this calculation we have used the vanishing of the Ricci tensor
in vacuum and some well known curvature tensor symmetry properties.

 Integrating over a
small coordinate sphere in the surface $x^0=0$, using
\begin{equation}
\int x^l x^m d^3x=\frac13\delta^{lm}\int r^2 d^3x,\qquad l,m=1,2,3
\end{equation}
and the traceless property of the Bel-Robinson tensor gives
\begin{equation}
P_\mu\simeq(2\kappa)^{-1}B^0{}_{\mu
lm}\delta^{lm}\frac{4\pi}{3\cdot4\cdot5}
r^5=(2\kappa)^{-1}B^0{}_{\mu 00}\frac{4\pi}{60}
r^5=\frac1{240G}B^0{}_{\mu 00}r^5.
\end{equation}
This result is best appreciated when expressed in terms of the
(traceless, symmetric) electric and magnetic parts of the Weyl
tensor, $E_{ab}:=R_{0a0b}$,
$H_{ab}:=\frac12\epsilon_{acd}R^{cd}{}_{0b}$. We then have a value
similar to that in electrodynamics:
\begin{equation}
P^{\mu}\simeq\frac{r^{5}}{240G}
(E_{ab}E^{ab}+H_{ab}H^{ab},-2\epsilon^{cab}E_{ad}H^{d}{}_{b});
\end{equation}
the most important feature is $P^0\ge|P^i|\ge0$.

\section{Conclusion}
Thus the desired Bel-Robinson property is {\it naturally\/}
satisfied for M{\o}ller's energy-momentum density. An important
consequence is that the gravitational energy according to this
measure is {\it positive\/}, at least to this order. (We expected
this positivity result since in fact M{\o}ller's tetrad expression
has an associated positive energy proof \cite{Nes89}.)

We stress that the vacuum small region Bel-Robinson property is a
strong test capable of excluding many otherwise acceptable
expressions;  in particular {\it none\/} of the classical
pseudotensors satisfy this requirement (although certain quite
artificial combinations of them do \cite{DFS99,icga7so,So06}). Once
again M{\o}ller's 1961 tetrad energy-momentum ``tensor'' stands out
as one of the best descriptions for gravitational energy-momentum.

%%%%%%%%%%%%%%%%%%%%%%%%%%%%%%%%%%%%%%%%%%%%%%%%%%%%%%%%%%%%%%%%%%%%%%%%%%%%%%%
\section*{Acknowledgments}
We would like to thank C. M. Chen  and the NCTS gravity and
cosmology focus group for many stimulating discussions as well as
the Taiwan NSC for their financial support under the grant numbers
NSC 93-2112-M-008-001, 94-2119-M-002-001, 94-2112-M008-038, and
95-2119-M008-27.

\end{document}